\documentclass[prl,twocolumn, showkeys]{revtex4}
\usepackage{amssymb, amsmath,  amscd, graphicx, color}
\newtheorem{theorem}{Theorem}
\newcommand{\champa}{\lambda}
\newcommand{\champb}{\lambda}
\newcommand{\champc}{\lambda}
\newcommand{\prestr}{{r_{\gamma}}}
\newcommand{\preste}{{q_\gamma}}

\begin{document}
\title{Rigorous Analysis of Singularities and 
Absence of Analytic Continuation\\ 
at First Order Phase Transition Points in Lattice Spin
Models~\footnote{Supported by the Swiss National Foundation for Scientific
Research.}}
\author{Sacha Friedli} 
\affiliation{Institute of Theoretical Physics, EPFL, 1015 Lausanne,
Switzerland}
\author{Charles-\'Edouard Pfister}
\affiliation{Institute of Analysis and Scientific Computing, School of
Mathematics, EPFL, 1015 Lausanne, Switzerland}
\begin{abstract}
We report about two new rigorous results on the non-analytic properties of
thermodynamic potentials at first order phase transition. The first one is valid
for lattice models ($d\geq 2$) with
arbitrary finite state space, and finite-range interactions
which have two ground states. Under the only assumption that the
Peierls Condition is
satisfied for the ground states and that the temperature is sufficiently low, 
we prove that the pressure has no analytic
continuation at the first order phase transition point. 
The second result 
concerns Ising spins with Kac
potentials $J_\gamma(x)=\gamma^d\varphi(\gamma x)$, where $0<\gamma<1$ is 
a small
scaling parameter, and $\varphi$ a fixed finite range potential. In this
framework, we relate the non-analytic behaviour of the pressure at the
transition point to the range of interaction,
which equals $\gamma^{-1}$.
Our analysis exhibits a
crossover between the non-analytic behaviour of finite range models
($\gamma>0$) and analyticity in the mean field limit ($\gamma\searrow 0$).
In general, the basic mechanism
responsible for the appearance of a singularity blocking the
analytic continuation is that
arbitrarily large droplets of the other phase become stable at the transition
point.

\end{abstract}
\maketitle
\section{Introduction}
The first theory of condensation originated with the celebrated equation of
state of van der Waals~\cite{vdW}:
\begin{equation}\label{E1}
\Big(p+\frac{a}{v^2}\Big)\big(v-b\big)=RT\,.
\end{equation} 
When complemented with the
Maxwell Construction (or ``equal area rule''), \eqref{E1} leads to isotherms
describing general characteristics of the liquid-vapor equilibrium, including
the existence of a critical temperature. The isotherms obtained with the 
van der Waals-Maxwell Theory have a very simple analytic structure: they are
analytic in a pure phase and have analytic continuations along the liquid and
gas branches, through the transition points. These analytic continuations, which
were originally interpreted as describing the pressure of metastable states, are
provided by the original isotherm given in \eqref{E1}.\\

The theoretical question of knowing
whether the results predicted by the van der Waals
Theory can be derived from first principles of Statistical Mechanics remained a
longstanding problem during a large part of the twentieth century. The 
theories of Mayer \cite{Mayer} and 
Yang-Lee \cite{YL}
were decisive contributions to the theory of phase transitions, 
but didn't give an answer concerning the delicate 
question
of the analytic continuation at transition points. With regard to this
latter property, two scenarios were discussed in the fifties and sixties.\\

The first one was essentially 
based on the mean field (or Bragg-Williams)
approximation \cite{TempKat}.
In this approach, the interaction
is replaced by an infinite range and infinitely weak potential.
The central characteristic of the effective model obtained after
this approximation is that the spatial positions of the
particles don't play any role.
As a consequence, an exact computation of the partition function leads to the
same behaviour as in 
the van der Waals-Maxwell Theory: at low temperature,
thermodynamic potentials are analytic in a pure phase, and have analytic
continuation at transition points.
 Katsura \cite{Kat} conjectured that this
scenario holds also for short range models, like the Ising model (see also the
discussion below).\\

The second argument, totally different in spirit,
originated with the so called ``droplet mechanism''
of the condensation phenomenon, proposed by Andreev \cite{Andreev},
Fisher \cite{Fisher} and Langer \cite{Langer}. 
This mechanism, as opposed to the mean field approximation,
predicts that the {finiteness} of the range of interaction plays a crucial
role in the analytic properties of the thermodynamic potentials. 
Namely, when the range of interaction is finite,  
droplets of any size are stable at the condensation point, and although the
probability of occurence of large droplets is very small, they yield a
contribution of the order $k!^{\frac{d}{d-1}}$ to the $k$-th derivative of the
pressure. Kunz and Souillard were led to
the same conclusions after having studied a similar model, related
to percolation \cite{KunzSou}.\\

Subsequent papers on the subject, in which no definite answer was given,
include \cite{NewmSchulm}, \cite{EntBaxt}, 
\cite{PrivmSchulm}. 
More recent studies can also be found in 
\cite{Gunter}, \cite{Penrose}, \cite{Meunier}.

\section{Rigorous Results}
The first rigorous result was the study of 
Isakov \cite{Isakov1} on the Ising 
model, which confirmed the droplet predictions:\\

\noindent{\bf Isakov (1984):}\emph{
In dimension $d\geq 2$, at low enough temperature,
the pressure of the Ising model
in a magnetic field $\champc$, $p=p(\champc)$, is infinitely differentiable at 
$\champc=0^\pm$, but has no analytic continuation from $\{\champc< 0\}$ to 
$\{\champc> 0\}$ accross $\champc=0$, or vice versa.}\\

This result is obtained by showing that the derivatives of 
the pressure at $\champc=0$ behave like
\begin{equation}\label{E3}
p^{(k)}(0^\pm)\sim C^kk!^{\frac{d}{d-1}}\,.
\end{equation}
In a second paper \cite{Isakov2}, Isakov
tried to
extend this result to general two phase lattice models. He had, however, to
introduce hypothesis that are not easy to verify in concrete models. We now
present our results.\\

\begin{figure}[htbp]
\includegraphics{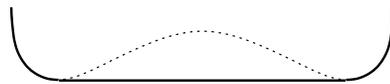}
\caption{The free energy in the van der Waals Limit.}
\label{F1}
\end{figure}

\paragraph{Two Phase Models.}
Consider a lattice model with finite state space at each site of
$\mathbb{Z}^d$, $d\geq 2$.
Let $H_0$ be a hamiltonian with finite range periodic interaction, having
two periodic ground states $\psi_1, \psi_2$, so that the Peierls 
Condition is satisfied \cite{PeierlsCond}.
Let $V$ be a periodic potential with finite range
interaction, so that the perturbed hamiltonian
\begin{equation}\label{E4}
H_\champa=H_0+\champa V
\end{equation}
splits the degeneracy of $H_0$. That is, $H_\champa$ has a single ground state
$\psi_2$ when $\champa<0$ and a 
single ground state $\psi_1$ when $\champb>0$. Denote
by $p=p(\champa)$ the pressure of the model (at inverse
temperature $\beta$).
Let $\delta>0$.
The general theory of Pirogov-Sinai \cite{PirogovSinai} 
guarantees that
if $\beta$ is large enough, then there exists $\champa^*(\beta)\in
(-\delta,+\delta)$ such
that the pressure has a first order phase transition at $\champa^*(\beta)$.
Our first 
result \cite{nous1} is the following:
\begin{theorem}\label{T2} There exists $\beta_0>0$ such that for all $\beta\geq
\beta_0$, the pressure
is analytic in $\champa$ on $(-\delta,\champa^*(\beta))$ and 
$(\champa^*(\beta),+\delta)$, but has no analytic continuation from
$(-\delta,\champa^*(\beta))$ to
$(\champa^*(\beta),+\delta)$ across $\champa^*(\beta)$ or vice-versa.
\end{theorem}
\paragraph{Kac Potentials and the van der Waals Limit.}
Consider an Ising ferromagnet, with a spin $\sigma_i\in\{+1,-1\}$ at each site
of $\mathbb{Z}^d$, $d\geq 2$. Let
$\varphi:\mathbb{R}^d\to \mathbb{R}^+$, supported by the cube $[-1,+1]^d$,
such that 
\begin{equation}\label{E5}
\int \varphi(x)\mathrm{d}x=1\,.
\end{equation}
Let $0<\gamma<1$ be a small scaling parameter, and consider the Kac potential
$J_\gamma(x)=\gamma^d\varphi(\gamma x)$,
together with the hamiltonian
\begin{equation}\label{E7}
H=-\frac{1}{2}\sum_{i\neq j}J_{\gamma}(i-j)\sigma_i\sigma_j\,.
\end{equation}
Let $f_\gamma=f_\gamma(m)$ denote the free energy of
this model, with fixed magnetization $m\in[-1,+1]$.
The Theorem of Lebowitz-Penrose \cite{LebPenrose}
gives a closed form to the free energy in the van der Waals limit 
$\gamma\searrow 0$ (called
sometimes the Kac or mean field limit),
and justifies the Maxwell construction.
Let $f_0(m)=\lim_{\gamma\searrow 0}f_\gamma(m)$. Then (see Figure \ref{F1})
\begin{equation}\label{E10}
f_0(m)=\mathrm{convex\,\, envelope\,\, of}\,
\Big\{-\frac{1}{2}m^2-\frac{1}{\beta}I(m)\Big\}\,,
\end{equation}
where $I(m)$ equals
\begin{equation}
I(m)=-\frac{1-m}{2}\log\frac{1-m}{2}-\frac{1+m}{2}\log\frac{1+m}{2}\,.
\end{equation}

\noindent When $\beta>1$, $f_0(m)$ has a plateau
$[-m^*(\beta),+m^*(\beta)]$, where
$m^*(\beta)$ is the positive solution of the mean field equation $m=\tanh(\beta
m)$. As a consequence of the Lebowitz-Penrose Theorem, all 
the analytic properties of the free
energy are known explicitly after the van der Waals limit: 
$f_0$ is analytic on the
branches $(-1,-m^*(\beta))$ and $(+m^*(\beta),+1)$, 
and has analytic continuation along
the paths $m\nearrow -m^*(\beta)$, $m\searrow +m^*(\beta)$. The analytic
continuation, which is unique, is given by the mean field free energy 
$-\frac{1}{2}m^2-\frac{1}{\beta}I(m)$. {After} the van der Waals limit, 
the scenario is thus the same as in the
van der Waals-Maxwell Theory. \\

Consider the specific choice
$\varphi(x)=2^{-d}1(x)$, where $1(\cdot)$
is the indicator of the cube: $1(x)=1$ if 
$x\in [-1,+1]^d$, $0$ otherwise.
For a fixed 
$0<\gamma<1$, $J_\gamma$ is finite range and
Theorem \ref{T2} can be used, but only for temperatures $\beta\geq
\beta_0(\gamma)$, with $\lim_{\gamma\searrow 0}\beta_0(\gamma)=+\infty$. 
Our result \cite{nous2} is given
hereafter. It holds at low temperature, uniformly in the range of
interaction.
\begin{theorem}\label{T3}
There exists $\beta_0$, independent of $\gamma$, 
such that for all $\beta\geq \beta_0$
and all $0<\gamma<1$, the free energy $f_\gamma$ is analytic on
$(-1,-m^*(\beta,\gamma))$ and $(+m^*(\beta,\gamma),+1)$, but has no 
analytic continuation along the real
paths $m\nearrow -m^*(\beta,\gamma)$, $m\searrow +m^*(\beta,\gamma)$.
\end{theorem}
This result shows that, as opposed to the mean field behaviour,
finite range interactions, even of very long range, imply
absence of analytic continuation at transition points. 
A crucial ingredient for the proof of Theorem \ref{T3}
is the use of a coarse-graining technique proposed by Bovier and Zahradn\' \i
k \cite{BovZarh}; it
allows to obtain a range of temperature that is uniform in $\gamma$.\\

\begin{figure}[hbp]
\begin{picture}(0,0)%
\includegraphics{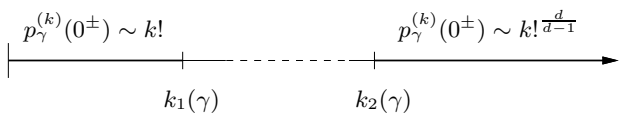}%
\end{picture}%
\setlength{\unitlength}{2881sp}%
\begingroup\makeatletter\ifx\SetFigFont\undefined%
\gdef\SetFigFont#1#2#3#4#5{%
  \reset@font\fontsize{#1}{#2pt}%
  \fontfamily{#3}\fontseries{#4}\fontshape{#5}%
  \selectfont}%
\fi\endgroup%
\begin{picture}(5294,814)(429,-944)
\put(3826,-286){\makebox(0,0)[lb]{\smash{\SetFigFont{9}{10.8}{\rmdefault}{\mddefault}{\updefault}{\color[rgb]{0,0,0}$p^{(k)}_\gamma(0^\pm)\sim k!^{\frac{d}{d-1}}$}%
}}}
\put(1801,-886){\makebox(0,0)[lb]{\smash{\SetFigFont{9}{10.8}{\rmdefault}{\mddefault}{\updefault}{\color[rgb]{0,0,0}$k_1(\gamma)$}%
}}}
\put(3451,-886){\makebox(0,0)[lb]{\smash{\SetFigFont{9}{10.8}{\rmdefault}{\mddefault}{\updefault}{\color[rgb]{0,0,0}$k_2(\gamma)$}%
}}}
\put(601,-286){\makebox(0,0)[lb]{\smash{\SetFigFont{9}{10.8}{\rmdefault}{\mddefault}{\updefault}{\color[rgb]{0,0,0}$p^{(k)}_\gamma(0^\pm)\sim k!$}%
}}}
\end{picture}
\caption{The crossover in the derivatives of the pressure.}
\label{F2}
\end{figure}

We study the pressure $p_\gamma=p_\gamma(\champb)$, in
which the constraint on the magnetization is replaced by a magnetic field 
$\champb$. The pressure and free energy are
related by a Legendre transform:
\begin{equation}
f_\gamma(m)=\sup_{\champb}\big(hm-p_\gamma(\champb)\big)\,.
\end{equation}
By the Theorem of Yang-Lee, $p_\gamma$ is analytic in $\champb$ on
$\{\champb<0\}$ and $\{\champb>0\}$.
Our main result is a precise
characterization of the properties of $p_\gamma$ along the path
$\champb\searrow 0$ (using symmetry, we need only consider fields $\champb> 0$).
\begin{theorem}\label{T4}
There exists $\beta_0$, independent of $\gamma$,
such that for all $\beta\geq \beta_0$
and all $\gamma>0$, all the limits $p_\gamma^{(k)}(0^+)=\lim_{\champb\searrow
0}p_\gamma^{(k)}(\champb)$ exist, but
pressure has no analytic continuation from 
$\{\champb>0\}$ to $\{\champb<0\}$
accross $\champb=0$. More precisely, there
exists integers $k_1(\gamma)$, $k_2(\gamma)$, 
$k_1(\gamma)<k_2(\gamma)$, with
$\lim_{\gamma\searrow 0}k_i(\gamma)=+\infty$,
such that
\begin{align}
|p_\gamma^{(k)}(0^+)|&\leq C_1^k k!&\text{ when }k\leq k_1(\gamma)\,,\\
|p_\gamma^{(k)}(0^+)|&\geq C_2^k k!^{\frac{d}{d-1}}&\text{ when }k\geq
k_2(\gamma)\,.
\end{align}
The constant $C_1$ is independent of $\gamma$ and $k$, 
$C_2=C_2(\gamma,\beta)>0$, and $k_1(\gamma)=\gamma^{-d}$. 
\end{theorem}
That is, the large order derivatives reveal the non-analytic feature of the
singularity, although a signature of the mean field (analytic)
behaviour can be detected in the low order derivatives.
We have illustrated this crossover on Figure \ref{F2}.

\section{Method}
The pressure has a singularity only in the thermodynamic limit. However,
we study the system in large finite volumes, and obtain bounds on the 
derivatives that are uniform in the volume.
At the end we prove that it is possible to interchange the operations of
taking the derivative and the thermodynamic limit.\\

The method used to obtain lower bounds on the derivatives of the pressure at
finite volume is inspired by the technique of Isakov.
Let $\Lambda$ be a finite cube in $\mathbb{Z}^d$ with a fixed boundary
condition, and $Z(\Lambda)$ be the
corresponding partition function. One enumerates all possible
contours \footnote{Contours 
are connected components of incorrect points (see 
 \cite{PeierlsCond}).} inside
$\Lambda$: $\Gamma_1,\Gamma_2,\dots,\Gamma_n$, in such a way that
$V(\Gamma_i)\leq V(\Gamma_j)$ when $i\leq j$ ($V(\Gamma_i)$ denotes the volume
of the interior
of the contour $\Gamma_i$). One then defines the restricted partition functions
$Z_i(\Lambda)$, $i=0,\dots,n$. By definition, $Z_0(\Lambda)$ is the partition
function computed for a system containing no contours, and $Z_i(\Lambda)$ is the
partition function computed for a system containing no contour 
$\Gamma_j$ with $j>i$. Obviously,
\begin{equation}\label{E22}
Z(\Lambda)=Z_0(\Lambda)\prod_{i=1}^n\frac{Z_i(\Lambda)}{Z_{i-1}(\Lambda)}\,.
\end{equation}
For the proof of Theorem \ref{T2}, there is only the ground state configuration
contributing to $Z_0(\Lambda)$. For the proof of Theorem \ref{T3},
$Z_0(\Lambda)$ is the partition function of a restricted
phase, describing small local fluctuations of the ground state. Let
\begin{equation}
u_\Lambda(\Gamma_i)=\log\frac{Z_i(\Lambda)}{Z_{i-1}(\Lambda)}\,.
\end{equation}
Notice that we have the fundamental relation 
\begin{equation}
Z_i(\Lambda)
=Z_{i-1}(\Lambda)+Z_{i-1}^*(\Lambda)\,, 
\end{equation}
where the contour $\Gamma_i$ appears in each configuration contributing to 
$Z_{i-1}^*(\Lambda)$. 
A precise analysis of the phase diagram shows that
$\champa\mapsto u_\Lambda(\Gamma_i)(\champa)$ is
analytic in a disc $\mathcal{U}_{i}$ centered at
$\champa=\champa^*(\beta)$ (resp. $\champb=0$ for the Kac ferromagnet), 
with a radius of order $V(\Gamma_i)^{-\frac{1}{d}}$. In the domain 
$\mathcal{U}_i$,
$u_\Lambda(\Gamma_i)$ can be represented as follows:
\begin{equation}
u_\Lambda(\Gamma_i)
=\log\Big(1+\frac{Z_{i-1}^*(\Lambda)}{Z_{i-1}(\Lambda)}\Big)
\equiv\log(1+e^{g_{\Lambda}(\Gamma_i)})\,.
\end{equation}
The dependence of $g_{\Lambda}(\Gamma_i)$ on the volume $\Lambda$ is weak. 
Moreover,
${g_{\Lambda}(\Gamma_i)}$ can be decomposed into a surface term and a
volume term, like in the droplet model.
Then, by choosing a path of integration $\mathcal{C}\subset\mathcal{U}_i$,
\begin{equation}\label{E15}
u_i(\Lambda)^{(k)}(\champa^*)=\frac{k!}{2\pi
i}\int_{\mathcal{C}}\frac{u_i(\Lambda)(\champa)}{(\champa-\champa^*)^{k
+1}}\mathrm{d}\champa\,.
\end{equation}
The observation of Isakov is that $u_\Lambda(\Gamma_i)\simeq 
e^{g_\Lambda(\Gamma_i)}$ on $\mathcal{U}_i$, and that
for a given large enough $k$, one can estimate precisely
the Cauchy integral \eqref{E15}, for all large enough contours, by a stationary
phase method, choosing suitably the path of integration $\mathcal{C}$. In this
way one gets a contribution to the $k$-th derivative of the pressure of order
$A^kk!^{\frac{d}{d-1}}$. For the other contours, only an upper bound can be
obtained on the integral, of the same 
order $B^kk!^{\frac{d}{d-1}}$. The crucial
point is therefore to have large enough neighbourhoods 
$\mathcal{U}_{i}$, in order to show that $A>B$. 
\section{Discussion}
In the framework of Kac potentials, the role played by the range of interaction
in the analyticity properties of the pressure
can be clarified by the following discussion. When $\champb\geq 0$, 
our analysis allows to decompose the pressure in two distinct parts: 
\begin{equation}\label{E32}
p_\gamma=\prestr+\preste\,.
\end{equation}

On one hand, $\prestr$ is constructed with the partition function $Z_0(\Lambda)$
of \eqref{E22}, and describes a homogeneous phase with positive
magnetization, containing no droplets of the $-$ phase. 
When $\gamma\searrow 0$, $\prestr$ converges to the pressure of the mean field
model.
On the other hand,
$\preste$ contains the contributions from the droplets of the $-$ phase,
which are all stable at $\champb=0$, and 
$\preste=O(e^{-\beta\gamma^{-d}})$. Namely, 
the main contribution to $\preste$ comes from 
the smallest droplets, which live on a
coarse-grained lattice whose cells have side length $\gamma^{-1}$.
Then, the pressure
$\prestr$ behaves analytically at $\champb=0$, i.e.
$\prestr^{(k)}(0^\pm)\sim k!$ for all $k$, but $\preste$
is
responsible for the absence of analytic continuation at $\champb=0$, since
$\preste^{(k)}(0^\pm)\sim k!^{\frac{d}{d-1}}$ for large enough $k$. The
combination of these two behaviours
leads to a crossover in the derivatives, as was shown in Theorem \ref{T4}.\\

Our results also have an important consequence regarding the theory of
condensation initiated by Mayer \cite{Mayer}. In this theory, the pressure of a
non-ideal gas is described,
near $z=0$ ($z$ is the fugacity), by a  
convergent Taylor expansion, given by the Mayer series: 
\begin{equation}\label{E30}
\beta p(z)=\sum_{l\geq 1}b_lz^l\,.
\end{equation}
(The $b_l$ are the cluster coefficients.)
That is, the pressure is analytic in the gas phase.
Then, the condensation point is defined to be the first singularity
encountered when \eqref{E30} is continued analytically along the positive real
line $z> 0$. It was suggested by 
some authors \cite{TempKat} that this method could
actually lead to a wrong determination of the condensation point: the analytic
continuation of the Mayer series 
might not ``see'' the real
transition point $z_c^*$, situated somewhere between $0$ and $z_M^*$:
$0<z_c^*<z_M^*$. 
This is indeed the case in the mean field approximation: the system does not
``see'' the condensation point, since there are no droplets of the liquid phase
inside the gaseous phase. We saw that if one suppresses the
condensation mechanism
by retaining, in \eqref{E32}, only the term $r_\gamma$,
then there is an analytic continuation of the pressure.
Our analysis shows that the phenomenon of condensation itself is responsible,
through the singular term $q_\gamma$,
for the singularity. Therefore, the latter method of determining the
condensation point gives the correct result,
at least for a large class of lattice gas models.\\

To conclude, we mention that the problem of knowing whether 
the pressure can be
continued analytically around the singularity, in the complex plane,  
remains open. It is not clear, in this case, whether the droplet
models can be used as a guiding mechanism, even to give a
heuristic description~\footnote{See the discussion in 
\cite{Penrose}, p.274, before Theorem 1.}.


\begin{thebibliography}{2000}
\bibitem{PeierlsCond}
Without restricting the generality we can assume that the energy 
$E_0(\psi_i)$ (associated to $H_0$) of each ground state $\psi_i$ is zero. 
Let $y\in\mathbb{Z}^d$, $l_0>R$ ($R$ denotes the range
of interaction), and $C_y=\{x\in\mathbb{Z}^d:|x_i-y_i|\leq l_0,\,
i=1,\dots,d\}$. Given a spin configuration $\omega$, which differs 
from a ground state
configuration only at finitely many points, we say that $y$ is correct if the
configuration $\omega$, restricted to $C_y$, coincides with one of the ground
states. Let $I(\omega)$ denote the set of points that are not correct.
The Peierls Condition means that there exists a constant
$\rho>0$ such that $E_0(\omega)\geq \rho|I(\omega)|$.
That is, regions of incorrect points are energy barriers.
\end{thebibliography}
\end{document}